**RESEARCH ARTICLE**  **OPEN ACCESS**

# A new deepening of mass-radius empirical relation for main sequence stars

Salvatore Camposeo*

Istituto Nazionale di Fisica Nucleare Sezione di Bari, Università di Padova, Politecnico di Bari, Italy

**Abstract.** In the following paper I have compared some typical mass-radius relations for main sequence stars by studying their level of agreement with DEBCat (J. Southworth, *ASP Conf. Ser.* **496** (2015), https://doi.org/10.48550/arXiv.1411.1219, https://www.astro.keele.ac.uk/jkt/debcat/debs.dat), which is a recent catalog by J. Southworth. Models chosen for testing were originally developed using older, smaller datasets than DEBCat. Each model follows a two-piece function structure, where each branch is a monomial power-law. This approach is motivated by theoretical considerations suggesting that low-mass and high-mass main sequence stars exhibit distinct behaviors in energy production and transport. Best level of agreement with DEBCat is found for Zamorano's model (J. Zamorano, A. Gimenez, *Astrophys. Space Sci.* **114** (1985), https://doi.org/10.1007/bf00653969). Also a new empirical relation is proposed by fitting a two-piece monomial power-law to DEBCat main sequence stars.

## 1 Introduction

Main sequence (MS) stars sustain nuclear fusion of hydrogen inside their core. Our Sun is an example of main sequence star.

Nuclear fusion of hydrogen $^1_1H$ to helium $^4_2He$ can be achieved through two different paths: pp-cycle or CNO cycle, depending on the core temperature. Inside our Sun pp-cycle prevails, but in hotter MS stars (which necessarily are more massive [1]) CNO cycle prevails.

Over the past 70 years, general features of MS stars have been widely studied. The relation between radius and mass is one of these features.

This relation is not only intrinsically interesting, but it's also extremely useful when you know only radius or only mass (the latter case is the most common) and you want to estimate the other quantity. Clearly this empirical relation can be "built" by exploiting stars whose mass and radius are both (independently) known.

Two-piece mass-radius functions are strongly suggested by theoretical studies.

In particular, current theoretical models predict that low-mass and high-mass MS stars have different heat-transport behaviors. The first ones should be internally radiative and externally convective, while the second ones are expected to be internally convective and externally radiative [2]. The Sun falls into the former category [2]; so the separation between two behaviors is expected to occur for a switch-mass $M_{sw.HT}$ which is bigger than 1 $M_\odot$, i.e. more than one solar mass.

Another duality of behaviors theoretically expected is that of nuclear fusion paths, which has already been mentioned. Indeed there is a transition from pp-cycle prevalence (low-mass stars) to CNO cycle prevalence (high-mass stars). The switch-mass value for this transition, $M_{sw.FP}$, is expected to coincide with $M_{sw.HT}$, but this topic is far beyond the scope of this paper. Let's simply say that, when CNO cycle starts to prevail over pp-cycle, radiative gradient drastically grows and (because of Schwarzschild criterion) convection is strongly triggered in the inner part of the star [1]. The two discussed switch-mass values are thought to be "located" between 1.2 $M_\odot$ [1] and 2.0 $M_\odot$ [3,4].

Given the previous theoretical predictions, best hypothesis for the mass-radius relation is clearly a two-piece function. The most important parameter for this type of function is the switch-mass $M_{sw.R}$. Link between $M_{sw.R}$, $M_{sw.HT}$ and $M_{sw.FP}$ won't be investigated in this paper but certainly we expect $M_{sw.R}$ to lie between 1.2 $M_\odot$ and 2.0 $M_\odot$ [4].

From now on, M (mass) and R (radius) of stars will be expressed in solar units $M_\odot$ and $R_\odot$.

A typical mass-radius (two-piece) function is the following one:

$$\begin{aligned} R &= 1.00\, M^{0.98} \quad if\ M < 2.0 \\ &= 1.25\, M^{0.64} \quad if\ M > 2.0. \end{aligned} \quad (1)$$

---

* e-mail: `salvatore.camposeo@phd.unipd.it`

Equation (1) is the Kopal's mass-radius relation for MS stars [5].

Another typical mass-radius function is the Lacy's one [6]:

$$\begin{aligned} R &= 0.95\, M^{0.92} \quad if\ M < 1.3 \\ &= 1.03\, M^{0.64} \quad if\ M > 1.3. \end{aligned} \quad (2)$$

A more recent relation is that of Zamorano [4]:

$$\begin{aligned} R &= (1.13 \pm 0.02)M^{0.98 \pm 0.03}\ if\ M < 1.8 \\ &= (1.42 \pm 0.10)M^{0.56 \pm 0.04}\ if\ M > 1.8. \end{aligned} \quad (3)$$

Showed equations are empirical (continuous) functions derived from old catalogues. Each of the given functions has two branches. Each branch is a simple monomial power-law.

Uncertainties on parameters are not available for Kopal's model and for Lacy's one, since authors didn't give them [5,6].

For Zamorano's model, uncertainties on parameters are taken from its source paper [4], with the exception of error on $M_{sw.R}$ which is not given by authors.

Let's note that three showed functions almost entirely cover the previously hypothesized range of $M_{sw.R}$ values.

In this paper I will refer to mass-radius equations also with the label "models" (example: Model 1 is Eq. (1)).

It's immediate to observe that Model 1 gives R = 1 for M = 1, while Model 2 slightly underestimates solar radius (–5%) and Model 3 slightly overestimates it (+13%).

Indeed Model 1 has been built by fixing a priori R(1) = 1, so it has one less free parameter with respect to Model 2 and 3 [5].

Model 1, Model 2 and Model 3 have been built with old catalogues, approximately coinciding with Popper's dataset [7].

I have found also another two-piece relation for radius of MS stars: the Demircan's model [8]. It is not considered in this paper because it is not a continuous function (see Appendix A). Let's remind that research of an empirical mass-radius relation is strongly motivated by computational reasons (usually only mass or only radius of a star is directly measured and other quantity has to be computed). Lack of continuity (at $M = M_{sw.R}$) makes Demircan's model inadequate for this purpose.

This paper also excludes all multi-piece models (with more than two branches, such as recent Eker's model based on DEBCat [9]). They are extremely useful for computational purposes but they do not fit our assumption of behavior-duality previously discussed.

No other two-piece continuous mass-radius functions (besides Model 1, 2 and 3) were found in the literature by the author of this paper.

## 2 Method

This study is based on the star population given by the DEBCat [10]. DEBCat is a recent and constantly updated catalog of well-studied detached eclipsing binaries. It is freely available online and its version of September 2024 will be considered here.

Let's specify that most of stars exploited by Kopal et al., Lacy et al., Zamorano et al. are included in DEBCat, which contains far more objects than Popper's catalog.

DEBCat inclusion criteria are the following [10]:

A) evolution of two stars of the system is currently unaffected by binarity;
B) both mass and radius have been experimentally measured for both components with an uncertainty below 2%;
C) spectrum of both components is well measured so that surface temperature (which is roughly equivalent to effective temperature $T_{eff}$) is computable by fitting a Plank function.

The author of DEBCat has computed masses through Kepler's third law (being orbital period and separation known).

Radius has always been computed by the author of the catalog through following considerations.

Luminosity of a star is given by Stefan- Boltzmann equation:

$$L = 4\pi R^2 \sigma T_{eff}^4 \quad (4)$$

where $\sigma$ is Stefan constant.

We can rewrite Equation (4) in this way:

$$4\pi d^2 \phi = 4\pi R^2 \sigma T_{eff}^4 \quad (5)$$

where $d$ is the distance between observer (Earth) and binary system, and is a known quantity thanks to parallax method. $T_{eff}$ is a known quantity too. $\phi$ is the integral observed flux, so it is a directly measured quantity. So, thanks to equation (5), R can be computed for each star of each system.

DEBCat gives also uncertainty for mass and radius of each star (always below 2% for both quantities). DEBCat contains almost 400 stars, but only 316 stars will be considered in this paper, because only them are classified as V luminosity class stars, which are expected to be MS stars [2]. Classification is given by the author of DEBCat and is based on spectral features.

The extracted population is represented in panels A and B of Figure 1. Its 316 stars have different spectral types (O, B, A, F, G, K, M) but they all belong to the V luminosity class.

Figure 1.A shows a clear main sequence shape without giant or super-giant branches.

At a later time, a more strongly filtered population will be considered.

Testing of selected models on the extracted DEBCat population is based on a simple reduced-chi-squared (RCS) test. Lower is RCS, better is the level of agreement between model and data.

During MS star evolution, radius is theoretically expected to slightly and progressively increase [2]. So specific age (within main sequence lifetime) is the biggest cause of dispersion in mass-radius scatter plot.

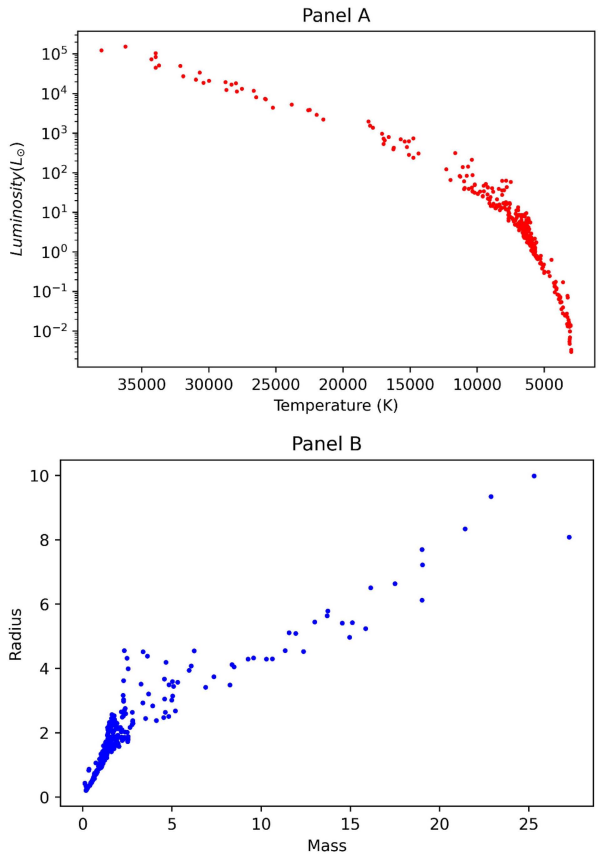

**Fig. 1.** Panel A: HR diagram of 316-star population extracted from DEBCat (both Temperature and Luminosity refer to photosphere). Panel B: mass-radius scatter plot of 316-star population (both quantities are expressed in solar units).

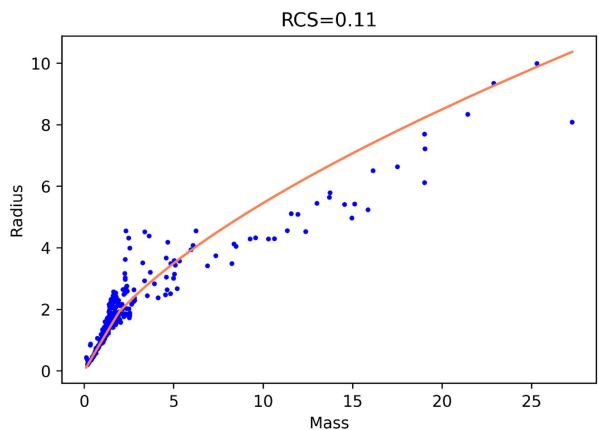

**Fig. 2.** Model 1 is represented by superimposing equation (1) on Figure 1b.

## 3 Results and discussion

In Figures 2, 3, 4, the three selected functions are superimposed on the mass-radius scatter plot of the 316-star population.

RCS has been computed for each model, considering the 316-star population:

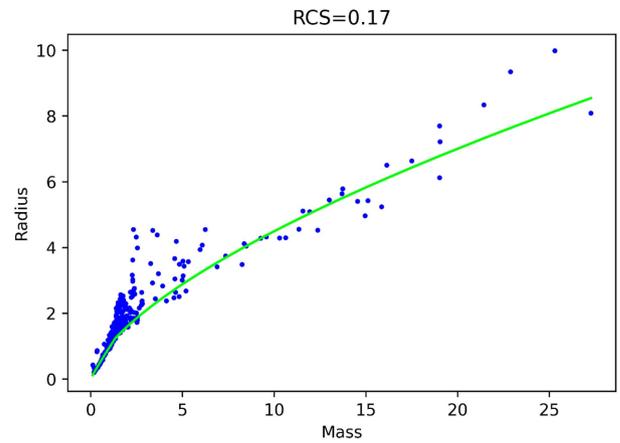

**Fig. 3.** Model 2 is represented by superimposing Equation (2) in Figure 1b.

RCS (Model 1) = 0.11
RCS (Model 2) = 0.17
RCS (Model 3) = 0.08.

This RCS compuation shows that Model 3 is the "favored one" (by DEBCat data).

Now I'm going to deepen the comparison between models by defining the "agreement band".

Let's consider the range going from

$$R_{Model}/A \text{ to } A \bullet R_{Model}$$

with $A > 1$, where $R_{Model}$ is the expected radius for a given mass, for a given model.

Let's call this range with the label "agreement band". It is possible to count (for each Model, for different values of $A$) the number of stars (among the 316-star population) whose measured radius does not enter the agreement band.

Lower is the number of stars excluded from the agreement band, better is the level of agreement between model and data.

In order to make this approach as least arbitrary as possible, I won't choose a specific fixed value of A. Comparison between models will be shown for many values of $A$ going from 1.2 to 2.5.

It would be unuseful to consider very small values of $A$ (closer to 1) because all stars tend to be excluded from agreement band when $A$ approaches 1 (see Fig. 5). Similarly, it would be unuseful to consider very big values of $A$ since all stars would be included in the agreement band (see Fig. 5).

Indeed when all stars are included or all stars are excluded (in/from the band), comparation of models is not possible.

In Figure 5 the number of excluded stars (from the agreement band) is plotted against the value of A for each model.

Most counts in Figure 5 have turned out to be "underdense" stars, i.e. stars whose measured radius is strongly bigger than model-predicted values.

It's immediate to observe that, with this approach, Model 2 is the worst model and Model 3 is the best one.

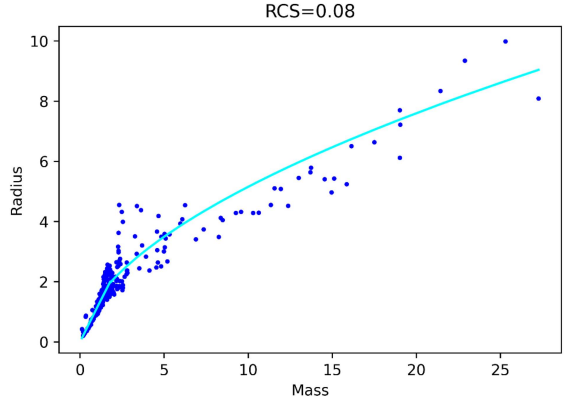

**Fig. 4.** Model 3 is represented by superimposing Equation (3) in Figure 1b.

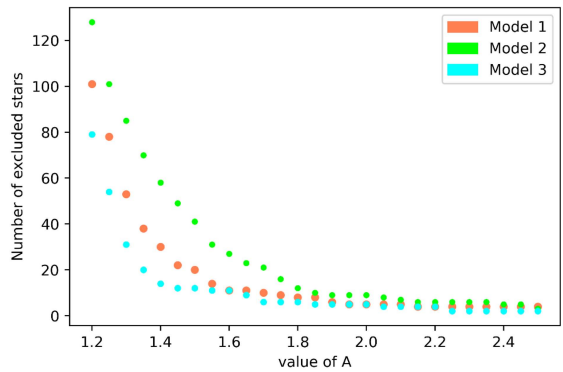

**Fig. 5.** Number of excluded stars VS value of A, for Model 1, 2 and 3, considering the 316-star population.

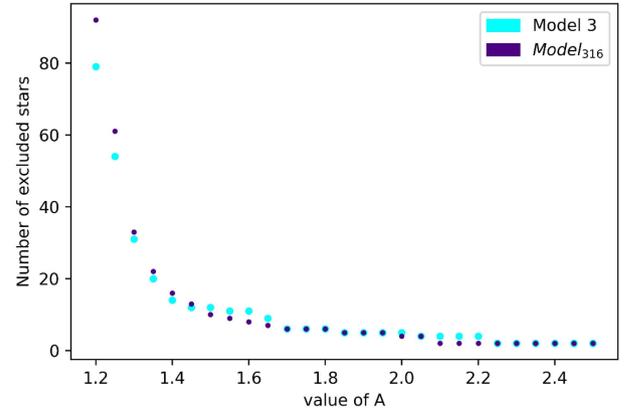

**Fig. 6.** Number of excluded stars VS value of A, for Model 3 and $Model_{316}$, considering the 316-star population.

So results of RCS test are fully confirmed.

An attempt has been made to compute a new two-piece mass-radius function by exploiting the 316-star population (imposing the usual branch type, i.e. a monomial power-law).

Imposed criteria are the following:

- continuity (to force at $M = M_{sw.R}$);

- 1.2 $M_\odot \leq M_{sw.R} \leq$ 2.0 $M_\odot$ (because of theoretical assumptions discussed in 'Introduction').

Fitting has been done through RCS-minimization method.

Resulting empirical model is given by the following function ('$Model_{316}$'):

$$R = (1.13 \pm 0.04)M^{0.92 \pm 0.06} \ if \ M < 1.99$$
$$= (1.49 \pm 0.05)M^{0.52 \pm 0.01} \ if \ M > 1.99. \quad (6)$$

Error on $M_{sw.R}$ parameter is 0.01.

RCS ($Model_{316}$) = 0.08, which is equal to RCS (Model 3).

A comparison between $Model_{316}$ and Model 3 is now proposed in Figure 6 through the agreement-band approach.

The two curves in Figure 6 are clearly overlapping as the two RCS are equal. This is not surprising since $Model_{316}$ and Model 3 have (the same functional form and) close parameters. So $Model_{316}$ and Model 3 show the same level of agreement with data.

Analysis performed in this paper until this point is now going to be repeated with a more strongly filtered population.

Indeed Figure 5 shows the presence of many peculiar stars (mostly "underdense" ones). Therefore it is possible that some pre-main sequence stars (PMS) are "contaminating" the 316-star population. PMS stars are objects whose core has not started nuclear fusion yet and their energy fuel is gravitational contraction [11]. They are low-density stars which can resemble MS stars (also in HR diagram) [11]. Also a few "overdense" stars contribute to counts of Figure 5. These stars have a measured radius which is quite smaller than predicted one. These objects may be stars which have undergone an accretion process in a distant past (after the beginning of MS phase) [12].

Now let's build a new population with a further filtering criterion (besides the belonging to the V luminosity class):

- average density of the star has to be higher than solar one if its mass is below 1 $M_\odot$, and has to be lower than solar one if its mass is above 1 $M_\odot$.

Indeed K-type and M-type MS stars are theoretically expected to be significantly denser than our Sun while O-type, B-type, A-type, F-type MS stars are expected to be less dense than our Sun [2,13]. Figure 7 (representing 316-star population) accredits this claim.

The new filtering has surely discarded the most peculiar stars (mostly underdense ones). A 300-star population is now found, corresponding to stars of top-left and bottom-right quadrants of Figure 7.

A population having less peculiar stars (than the previous one) is more suitable for our purposes; however, some of the discarded peculiar objects may still be MS stars.

RCS computation is repeated with this new (and smaller) set of MS stars:

RCS (Model 1) = 0.19
RCS (Model 2) = 0.40
RCS (Model 3) = 0.16.

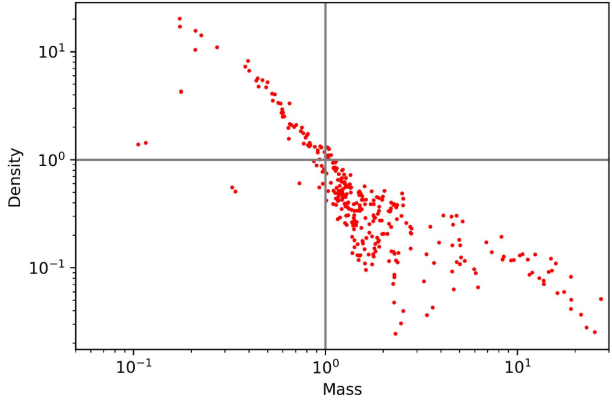

**Fig. 7.** Mass-density scatter plot of the 316-star population (logarithmic scale). Mass and (average) density are expressed in solar units. Grey lines correspond to solar mass and solar (average) density.

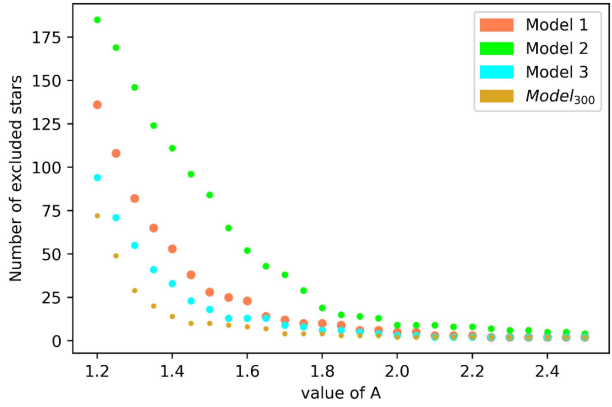

**Fig. 8.** Number of excluded stars VS value of A, for Model 1, 2, 3 and Model$_{300}$, considering the 300-star population.

Again Model 3 (i.e. Zamorano's function) is the one favored by RCS test.

Through RCS-minimization method, a new model is built again by fitting a two-piece monomial power law with the 300-star population. Again imposed criteria are:

– continuity (to force at $M = M_{sw.R}$);

– $1.2\,M_\odot \leq M_{sw.R} \leq 2.0\,M_\odot$.

Resulting function is the following, which I'm going to label as "Model$_{300}$":

$$R = (1.11 \pm 0.04)M^{0.98 \pm 0.07} \ if\ M < 1.86$$
$$= (1.48 \pm 0.05)M^{0.52 \pm 0.01} \ if\ M > 1.86. \quad (7)$$

Error on $M_{sw.R}$ parameter is 0.01.
In particular we find that:
RCS ($Model_{300}$) = 0.08.

A comparison between Model 1, Model 2, Model 3 and Model$_{300}$ is given in Figure 8 with the agreement-band approach, considering now the 300-star population.

**Table 1.** RCS values.

| 316-star pop. | Model 1: 0.11 | Model 2: 0.17 | Model 3: 0.08 | Model$_{316}$: 0.08 |
|---|---|---|---|---|
| 300-star pop. | Model 1: 0.19 | Model 2: 0.40 | Model 3: 0.16 | Model$_{300}$: 0.08 |

As suggested by RCS computation, in Figure 8 we see that Model$_{300}$ is glaringly favored over Model 1, Model 2 and Model 3.

## 4 Dead end

Observed mass-radius scatter plot (see Fig. 1B) has a strong dispersion because MS stars of our sample have different specific ages [2].

A model including age (within MS lifetime) could be hypotheiszed:

$$R = aM^\alpha \bullet f_{low}(s)\ if\ M < M_{sw.R}$$
$$= bM^\beta \bullet f_{high}(s)\ if\ M > M_{sw.R} \quad (8)$$

where $a$, $b$, $\alpha$ and $\beta$ are generic parameters; $f_{low}(s)$ and $f_{high}(s)$ are dimensionless functions; $s$ is the ratio between age (time passed since the beginning of Hydrogen burning) and expected (total) duration of MS lifetime (which is proportional to $M^{-2.5}$ [2]). Estimation of parameters of Equation (8) can't be performed because data about age is very poor and uncertain, in particular for stars above $1.8\,M_\odot$ [10].

A study involving age does exist but only for intermediate-mass stars (between $0.7\,M_\odot$ and $1.8\,M_\odot$) [14].

## 5 Conclusion

Relation between radius and mass of MS stars is not trivial: low-mass stars have an higher average density than high-mass stars.

In this paper three old two-piece mass-radius relations have been compared and tested on the recent star catalog DEBCat. Two distinguished populations of MS stars have been extracted from DEBCat by making two different filterings.

So the three models (Model 1, Model 2 and Model 3) have been tested and compared twice.

Also two new models have been formulated in this paper (one for each population):
Model$_{316}$ (Eq. (6)) and Model$_{300}$ (Eq. (7)) have been obtained by fitting a two-piece monomial power-law with the 316-star and with the 300-star populations respectively.

Testing has shown that Model 3 better fits DEBCat data with respect to Model 1 and Model 2 (for both 316-star and 300-star populations).

This paper strongly confirms the validity of Zamorano's model (i.e. Model 3), altough it is very old and built with a small number of stars (with respect to our two populations) [4]

Note that Model 3, Model$_{316}$ and Model$_{300}$ have <u>all parameters compatible within errors</u> except for M$_{sw.R}$, while Model 1 and Model 2 are strongly different.

Comparison between Model 3 and Model$_{316}$ (based on 316-star population) has not yielded a significant preference between the two; conversely Model$_{300}$ better fits data (of 300-star population) than Model 3 (see Tab. 1), <u>so that it can be considered an improved version of Zamorano's model</u>.

# Appendix A

Let's consider the following percentage:

$$P = \frac{|R_1(M_{sw.R}) - R_2(M_{sw.R})|}{0.5 \bullet [R_1(M_{sw.R}) + R_2(M_{sw.R})]} \bullet 100\% \quad (A.1)$$

where $R_1$ and $R_2$ indicate predicted radius by first and second branch respectively (for a given model).

Equation (A.1) quantifies the <u>discontinuity</u> of a model at $M = M_{sw.R}$.

Demircan's model is the following function [8]:

$$\begin{aligned} R &= 1.05\, M^{0.94} \quad if\ M < 1.66 \\ &= 1.37\, M^{0.54} \quad if\ M > 1.66. \end{aligned} \quad (A.2)$$

We find that:
P (Model 1) = 1%
P (Model 2) = 1%
P (Model 3) = 2%
P (Model$_{316}$) < 1%
P (Model$_{300}$) < 1%
P (Demircan) = 6%.

Since Demircan's <u>discontinuity percentage P</u> is almost an order of magnitude larger than others, the author of this paper has decided to discard this function.